\documentclass[12pt,a4paper,superscriptaddress]{revtex4}
\usepackage{graphicx}
\usepackage{graphics}
\usepackage{amsmath}
\usepackage{dcolumn}
\usepackage{amssymb}
\usepackage{bm}
\usepackage[latin1]{inputenc}
\newcommand{\be}{\begin{equation}}
\newcommand{\ee}{\end{equation}}
\newcommand{\bea}{\begin{eqnarray}}
\newcommand{\eea}{\end{eqnarray}}

\newcommand{\pa}{\partial}

\def\Bs{B\!\!\!/}

\def\Ds{D\!\!\!\!/}

\def\sech{\mathrm{sech}}

\begin{document}

\title{On the Adler-Bell-Jackiw anomaly in a Horava-Lifshitz-like QED}

\author{T. Mariz}
\affiliation{Instituto de F\'\i sica, Universidade Federal de Alagoas\\ 
57072-270, Macei\'o, Alagoas, Brazil}
\email{tmariz@fis.ufal.br}

\author{J. R. Nascimento}

\affiliation{Departamento de F\'{\i}sica, Universidade Federal da 
Para\'{\i}ba\\
 Caixa Postal 5008, 58051-970, Jo\~ao Pessoa, Para\'{\i}ba, Brazil}
\email{jroberto,petrov@fisica.ufpb.br}

\author{A. Yu. Petrov}

\affiliation{Departamento de F\'{\i}sica, Universidade Federal da 
Para\'{\i}ba\\
 Caixa Postal 5008, 58051-970, Jo\~ao Pessoa, Para\'{\i}ba, Brazil}
\email{jroberto,petrov@fisica.ufpb.br}

\begin{abstract}
We show the absence of the Adler-Bell-Jackiw (ABJ) anomaly for a Horava-Lifshitz-like QED with any even $z$.  Besides of this, we study the graph contributing to the ABJ anomaly at non-zero temperature and extend the Fujikawa's methodology of studying the integral measure for our model.
\end{abstract}
\maketitle

\section{Introduction}
During last years, the studies of field theory models with a strong space-time asymmetry, also known as Horava-Lifshitz-like (HL-like) models, attract a great attention. Originally such models were inspired by studies of critical phenomena many years ago \cite{Lif}. First studies of such models within the context of the quantum field theory were performed in \cite{Anselmi} where their renormalizability has been discussed. Further, formulation of the Horava-Lifshitz gravity \cite{Hor} crucially increased the interest to this class of models. Certainly, the HL-like gauge theories play a very important role within these models. The most interesting results for these studies are, first, obtaining the two- and three-point functions for five-dimensional \cite{Iengo} and four-dimensional \cite{our4d} cases, second, explicit calculation of the effective potential in the HL-like scalar QED \cite{EP}.

All this certainly calls the interest to study of more sophisticated aspects of the HL-like theories. One of them is just the problem of anomalies, especially the famous Adler-Bell-Jackiw (ABJ) anomaly (triangle anomaly) \cite{ABJ} implying breaking of the chiral symmetry. It is known that just this anomaly causes ambiguities in the theories with ``small'' Lorentz symmetry breaking \cite{JackAmb}. Therefore, it is natural to verify the presence of such anomaly in the HL-like extension of the QED. Namely this problem is considered in this paper.

\section{An HL-like Abelian gauge model}

Let us formulate the HL-like extension of QED involving coupling with the extra pseudovector field $B^{\mu}$.
For the sake of the concreteness and simplicity, we restrict ourselves to the case of even $z=2n$. The motivation for this study consists in the interest to study whether any Horava-Lifshitz-like analogue of the Carroll-Field-Jackiw term $\epsilon^{\mu\nu\rho\sigma}B_{\mu}A_{\nu}\partial_{\rho}A_{\sigma}$ known to be related with the ABJ anomaly in usual Lorentz-breaking theories \cite{JackAmb} can arise, and such a term involves a pseudovector $B_{\mu}$ as a necessary ingredient.

The results we obtained can be straightforwardly generated for the case of an arbitrary even critical exponent.
In this case, the Lagrangian of the spinor sector of the theory is
\bea
\label{arbz}
L&=&
\bar{\psi}(i\gamma^0D_0+(i\gamma^iD_i)^{2n}+m^{2n}+\Bs\gamma_5)\psi,
\eea
where $z=2n$ is a critical exponent. Here, the $D_{\mu}=\partial_{\mu}-ieA_{\mu}$, with $\mu=0,i$ and $i=1,2,3$,
is a gauge covariant derivative, with the corresponding gauge transformations being $\psi\to e^{ie\xi}\psi$, $\bar{\psi}\to \bar{\psi}e^{-ie\xi}$, and $A_{0,i}\to A_{0,i}+\pa_{0,i}\xi$, and $B_{\mu}$ is an extra pseudovector field. The dimensions of our objects look like follows: the dimension of $\partial_i$ and $eA_i$ is 1, as in the usual case, of $\partial_0$ and $eA_0$ is $z$, as the Horava-Lifshitz formalism requires, of $\psi$ is $d/2$, and of $B_{\mu}$ is $2n$. We note that in the absence of kinetic term for the vector field, the dimension of $A_0$ and $A_i$ fields cannot be fixed unambiguously. Also, we introduced $m^{2n}$ into the mass term for the spinor field in order to have the constant $m$ with a mass dimension equal to one.

The free propagator for the fermionic fields is
\bea
<\psi(k)\bar{\psi}(-k)>&=&S(k)=i\frac{\gamma^0k_0-(\vec{k}^{2n}+m^{2n})}{k^2_0-(\vec{k}^{2n}+m^{2n})^2}.
\eea

In principle, the number of different vertices for an arbitrary $n$ is very large. However, the ABJ anomaly involves as usual two vector fields and one axial field. If we consider the vertices with no more than two vector fields and no more than one derivative applying within any vertex, the number of vertices is drastically restricted, so, denoting $\nabla^2=\partial^i\partial_i$, we have only
\bea
\label{vertspin1}
V_1&=&ie\bar{\psi}\gamma^0A_0\psi,\quad\, V_2=ieC_2\bar{\psi}\gamma^i\gamma^j(\pa_iA_j)\nabla^{2n-2}\psi,\nonumber\\
V_3&=&2ieC_3\bar{\psi}A^i\pa_i\nabla^{2n-2}\psi, \quad\, V_4=e^2C_4\bar{\psi}A_iA^i\nabla^{2n-2}\psi.
\eea
In the momentum space they look like
\bea
V_1&=&ie\bar{\psi}(k)\gamma^0A_0(p)\psi(-p-k),\quad\,V_2=eC_{2,n}p_i\bar{\psi}(k)\gamma^i\gamma^jA_j(p)
(\vec{p}-\vec{k})^{2n-2}\psi(-p-k);\nonumber\\
V_3&=&-2C_{3,n}(p_i+k_i)e\bar{\psi}(k)A^i(p)(\vec{p}-\vec{k})^{2n-2}\psi(-p-k) ,\nonumber\\ V_4&=&e^2C_{4,n}\bar{\psi}(k_1)A_i(p_1)A^i(p_2)(\vec{k}_2)^{2n-2}\psi(k_2)(2\pi)^{d+1}\delta(k_1+k_2+p_1+p_2).
\eea
Here $C_{2,n},C_{3,n},C_{4,n}$ are the numbers generated by permutations of the Dirac matrices. Their explicit form is not important for us, however, it can be found, that is, $C_{2,n}=n(-1)^{n-1}$, $C_{3,2k}=2k$, $C_{3,2k+1}=-(2k+1)$,
$C_{4,2k}=2k$, $C_{4,2k+1}=-(2k+1)$.
There is also the extra vertex $\bar{\psi}\Bs\gamma_5\psi$.

\section{ABJ anomaly}

The ABJ anomaly is given by the Feynman diagram depicted at Fig.1.

%\vspace*{1mm}

\begin{figure}[h]
\includegraphics[scale=.8]{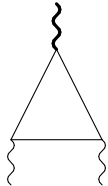}
\caption{General form of the contribution to the triangle anomaly.} \label{triangle}
\end{figure} 

%\vspace*{1mm}

Here the thick wavy line is for the external $B_{\mu}$ field. We note that the quartic vertex does not contribute to ABJ anomaly yielding a total derivative.

There are six possible contributions to it characterized by different positions of vertices $V_1$, $V_2$, $V_3$ (beside of the axial vertex). It is clear that the diagrams with two $V_1$ vertices and with two $V_2$ vertices do not contribute to the ABJ anomaly (indeed, in first of these cases the indices of external $A_0$ fields are the same, and in the second one, the contribution is of the second order in derivatives). The remaining four graphs can be described as follows. 
The first one involves $V_1$ and $V_2$ vertices, the second one -- $V_1$ and $V_3$ ones, for the third one -- $V_2$ and $V_3$ ones, finally, the fourth one is formed with use of two $V_3$ vertices. They are depicted at Fig. 2.

%\vspace*{1mm}

\begin{figure}[h]
\includegraphics[scale=.8]{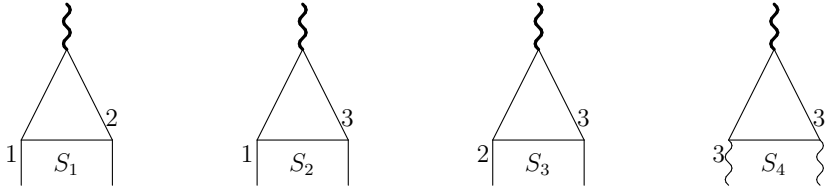}
\caption{Different contributions to the triangle anomaly.} \label{graphs}
\end{figure} 

%\vspace*{1mm}

The contributions of these graphs are
\bea
\label{contgraphs}
S_1&=&e^2C_{2,n}{\rm tr}\int\frac{dk_0d^3k}{(2\pi)^4}A_0(-p)\gamma^0S(k)\gamma^mB_m\gamma_5S(k)\gamma^i\gamma^j(\pa_iA_j)(p)S(k);\nonumber\\
S_2&=&ie^2C_{3,n}{\rm tr}\int\frac{dk_0d^3k}{(2\pi)^4}A_0(-p)\gamma^0S(k)\gamma^mB_m\gamma_5S(k)A^i(p)(k_i+p_i)S(k+p);\nonumber\\
S_3&=&4ie^2C_{2,n}C_{3,n}{\rm tr}\int\frac{dk_0d^3k}{(2\pi)^4}A^l(-p)k_lS(k)\gamma^mB_m\gamma_5S(k)\gamma^i\gamma^j(\pa_iA_j)(p)S(k);\nonumber\\
S_4&=&4e^2C_{3,n}^2{\rm tr}\int\frac{dk_0d^3k}{(2\pi)^4}A^l(-p)k_lS(k)\gamma^mB_m\gamma_5S(k)A^i(p)(k_i+p_i)S(k+p).
\eea
Here we omitted dependence of propagators on the external momentum $p$ within $S_1$ and $S_3$ since taking it into account will yield only the second- and higher-order contributions in derivatives.
It is clear that the contributions $S_2,S_3,S_4$ vanish. Indeed, in the $S_2$ and $S_4$ the structure of products of Dirac matrices is insufficient to give a non-zero trace, and the integrand of $S_3$ is odd with respect to the internal moments. 

So, it remains to consider $S_1$. Its explicit form is given by
\bea
S_1&=&e^2C_{2,n}{\rm tr}\int\frac{dk_0d^3k}{(2\pi)^4}A_0(-p)\gamma^0S(k)\gamma^mB_m\gamma_5S(k)\gamma^i\gamma^j(\pa_iA_j)(p)
{\vec{k}}^{2n-2}S(k).
\eea
By calculating the trace over the Dirac matrices, we obtain
\bea\label{S1}
S_1 = -4ie^2C_{2,n}\,\epsilon^{0ijk}A_0(-p)B_i A_j(p) p_k \int\frac{dk_0d^3k}{(2\pi)^{4}} \frac{[3k_0^2+(\vec k^{2n}+m^{2n})^2](\vec k^{2n}+m^{2n})\vec k^{2n-2}}{[k_0^2-(\vec k^{2n}+m^{2n})^2]^3}.
\eea
 A straightforward integration over $k_0$ shows that this integral vanishes for any $n$. We note that it is rather natural since the corresponding contribution is not gauge invariant.

\section{Finite temperature analysis of the ABJ anomaly}

Let us consider now the finite temperature study of the ABJ anomaly. To do it, we use the finite temperature formalism described in \cite{DJ} and generalized for the HL-like theories in \cite{highT}. General approach to chiral anomaly at the finite temperature can be found in \cite{Das}. Within our study we follow the lines of \cite{CSYM} where the triangle graph corresponding to the ABJ anomaly has been discussed at the finite temperature in the usual Lorentz-breaking QED. First, we carry out the Wick rotation:
\bea
\label{S1a}
S_1 = 4e^2C_{2n}\,\epsilon^{0ijk}A_0(-p)B_i A_j(p) p_k \int\frac{dk_0d^3k}{(2\pi)^{4}} \frac{[3k_0^2-(\vec k^{2n}+m^{2n})^2](\vec k^{2n}+m^{2n})\vec k^{2n-2}}{[k_0^2-(\vec k^{2n}+m^{2n})^2]^3}.
\eea
Let us now use the Matsubara formalism, which consists in taking $k_0=(n+\frac12)2\pi/\beta$ and changing $1/(2\pi)\int dk_0\to 1/\beta \sum$. Thus, we obtain
\be
S_1 = \frac{e^2}{m}\epsilon^{0ijk}A_0(-p)B_i A_j(p) p_k \int_0^\infty dK K^2 \frac{M}{\pi^3}\sum_n\frac{[3(n+\frac12)^2-(K^{2n}+M^{2n})^2](K^{2n}+M^{2n})}{[(n+\frac12)^2+(K^{2n}+M^{2n})^2]^3},
\ee
with $K^{2n}=\vec{k}^{2n}\frac{\beta}{2\pi}$ and $M^{2n}=m^{2n}\frac{\beta}{2\pi}$, where we have used spherical coordinates, i.e., $\int d^3k = (2\pi/\beta)^{\frac32}\int dK 4\pi K^2$.
Finally, one finds that Eq. (\ref{S1}) takes the form
\be
S_1 = e^2C_{2,n}m^{1-2n}\epsilon^{0ijk}A_0(-p)B_i A_j(p) p_k F(M)
\ee
with
\be
F(M) = \int_0^\infty dK K^{2n} M^{2n-1} \tanh[\pi(K^{2n}+M^{2n})]\sech^2[\pi(K^{2n}+M^{2n})].
\ee
It is easy to verify that the limits of high temperature and zero temperature in the function $F(M)$ vanish for any $n$. We note that disappearing of $F(M)$ at $T\to 0$ can be expected, first, by analogy with the case $n=1$, see \cite{our4d}, second, since we showed above that in the usual, zero temperature case the anomaly vanishes.

\begin{figure}[h]
\includegraphics[scale=.8]{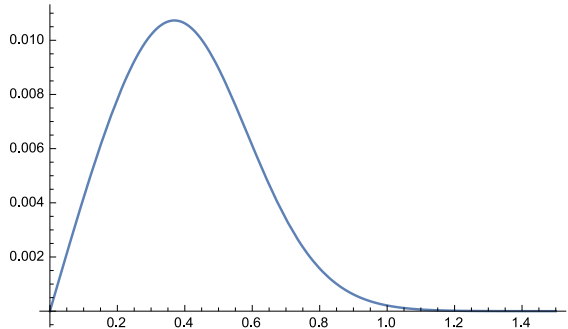}
\caption{Plot of the function $F(M)$} \label{Anomaly}
\end{figure}

The plot of the above function $F(M)$, which can be numerically calculated, is presented in Fig.~\ref{Anomaly}. Therefore, in the limit of high temperature ($T\to\infty$ or $M\to0$), or in the massless limit ($m\to 0$), the function $F(M)$ tends to zero, $F(M\to0)\to0$. Note that the limit of zero temperature ($T\to0$ or $M\to\infty$) is also confirmed, i.e., $F(M\to\infty)\to0$. Actually, restoring the explicit $T$ dependence of the function $F(M)$, one can show that at the high temperature, i.e. $\beta\to 0$, the leading term of the $F(M)$ is linear in $\sqrt{\beta}$. To verify this, one can expand $F(M)$ in power series in $M$ as
\bea
F(M)\simeq F(M=0)+\frac{dF}{dM}|_{M=0}M.
\eea
It is clear that $F(M=0)=0$. And
\bea
\frac{dF}{dM}|_{M=0}= \int_0^\infty dK K^{2n} \tanh[\pi K^{2n}]\sech^2[\pi K^{2n}],
\eea 
that is, a constant which we denote as $\rho$. Taking into account the explicit form of $M$, we find $F(M)\simeq 
\rho m^n\left(\frac{\beta}{2\pi}\right)^{1/2}$.

We conclude that the potential anomaly differs from zero only if temperature is neither zero nor infinite.

\section{Transformations of the measure}

It is well known that in the case of the usual QED \cite{Fuji} the chiral transformations of the spinor fields
\bea
\psi\to e^{i\alpha\gamma_5}\psi, \quad \bar{\psi}\to \bar{\psi}e^{i\alpha\gamma_5}
\eea
yield the additional contribution to the anomaly caused by the measure transformation and equal to
\bea
\delta S_{measure}=\frac{1}{16\pi^2}\int d^4x\alpha(x)\epsilon^{abcd}F_{ab}F_{cd}.
\eea
Let us consider the possibility of the analogous contribution in our case.

Instead of the base of fields $\phi_l$ satisfying the equation $\Ds\phi_l=\lambda_l\phi_l$ (remind that the usual spinor action is $\int d^4x\bar{\psi}(i\Ds-m)\psi$), in our case, we will have the base $\tilde{\phi}_l$ satisfying the equation $(i\gamma^0D_0-\Ds^{2n})\tilde{\phi}_l=\tilde{\lambda}_l\tilde{\phi}_l$, cf. (\ref{arbz}). Since both the operator $i\gamma^0D_0$ and the operator $\Ds^2$ are Hermitian, the operator $i\gamma^0D_0-\Ds^{2n}$ will also be Hermitian possessing thus only real eigenvalues. Therefore, all $\lambda_l$ are real.

The corresponding Jacobian will look like (cf. \cite{Fuji}):
\bea
\exp(-2iJ)=\exp(-2i\lim\limits_{N\to\infty}\sum\limits_{l=1}^{N}\int dtd^3x\alpha(t,x)\tilde{\phi}^{\dagger}_l(t,x)\gamma_5\tilde{\phi}_l(x)).
\eea
To provide the convergence of the integral, we introduce the regularization by inserting the function $f(\frac{\tilde{\lambda}^2_l}{M^4_0})$, where $M_0$ is a constant with a dimension of mass, so that $f(x)|_{x\to\infty}=0$. After proceeding as in \cite{Fuji}, the factor $J$ determining the Jacobian takes the form
\bea
J={\rm Tr}\int dtd^3x\alpha(t,x)\gamma_5f\left(\frac{(i\gamma^0D_0-\Ds^{2n})^2}{M^4_0}\right),
\eea
where, unlike \cite{Fuji}, $\Ds=\gamma^iD_i$ is a purely spatial contraction. The explicit form of the trace is
\bea
J={\rm Tr}\int dtd^3x\alpha(t,x)e^{-ikx}\gamma_5f\left(\frac{(i\gamma^0D_0-\Ds^{2n})^2}{M^4_0}\right)e^{ikx}.
\eea
One can argue as in \cite{Fuji} that only the second order in $(i\gamma^0D_0-\Ds^{2n})$ contributes to the measure (indeed, only it yields the quadratic contribution to the effective action), so, expanding the function $f$ up to the second order, we have
\bea
J=\frac{1}{2}{\rm Tr}\int dtd^3x\alpha(t,x)e^{-ikx}f^{\prime\prime}(0)\gamma_5\left(\frac{i\gamma^0D_0-\Ds^{2n}}{M^2_0}\right)^2e^{ikx}.
\eea
Then, the simple transformation shows that $(i\gamma^0D_0-\Ds^{2n})^2=(i\gamma^0D_0-(D^2+\frac{1}{4}[\gamma^i,\gamma^j]F_{ij})^n)^2$, thus,
\bea
J=\frac{1}{2}{\rm Tr}\int dtd^3x\alpha(t,x)e^{-ikx}f^{\prime\prime}(0)\gamma_5\left(\frac{i\gamma^0D_0-D^2+\frac{1}{4}[\gamma^i,\gamma^j]F_{ij}
}{M^2_0}\right)^2e^{ikx}.
\eea
However, unlike the usual case where the trace yields the Levi-Civita symbol, this expression involves either the trace of the product $\gamma_5[\gamma^i,\gamma^j][\gamma^k,\gamma^l]$ which is zero, or the trace of $\gamma_5\gamma^0[\gamma^i,\gamma^j]$ which is also zero. Hence, the factor $J$ is zero, and the Jacobian is consequently trivial being equal to 1. We conclude that in our case both the variation of the measure and the triangle contribution vanish.

\section{Summary}

Let us discuss our results. We find that the only nontrivial contribution vanishes at the zero temperature and tends to zero at a very high temperature. In principle, the vanishing of the CFJ-like contribution at the zero temperature has a natural reason -- actually, since the action of the theory does not involve any terms linear in the ``spatial'' Dirac matrices $\gamma_i$, it is not invariant under the chiral transformations already at the classical level. We explicitly demonstrated that the similar situation will occur for other even $z$ (indeed, in these cases the propagators and the vertices will involve even numbers of the spatial $\gamma^i$ Dirac matrices). Also, we have showed that the contribution to the effective action generated by the transformation of the measure is trivial.
As a by-product, we can see that the one-loop contribution to the pion decay (which can be obtained by replacement of $\Bs\gamma_5$ by $\gamma_5$ in the corresponding vertex) is zero since the number of space-like $\gamma_i$ matrices in its contribution is insufficient. 

All our discussions were carried out for the even $z$.
At the same time, the odd $z$ (we note that the contribution of integral measure to the possible chiral anomaly has been discussed in \cite{Bakas}, for the case of an essentially odd critical exponent $z$) will essentially differ. We are going to consider this situation in our next paper.

{\bf Acknowledgements.} This work was partially supported by Conselho
Nacional de Desenvolvimento Cient\'{\i}fico e Tecnol\'{o}gico (CNPq). The work by A. Yu. P. has been partially supported by the
CNPq project No. 303438/2012-6.


\begin{thebibliography}{99}
\bibitem{Lif} E. M. Lifshitz, Zh. Eksp. Teor. Fiz., 11, 255 \& 269 (1941). 
\bibitem{Anselmi} D. Anselmi, Ann. Phys. 324, 874 (2009), arXiv: 0808.3470; Ann. Phys. 324, 1058 (2009), arXiv: 0808.3474; R. Iengo, J. Russo, M. Serone, JHEP 0911, 020 (2009), arXiv: 0906.3477; P. R. S. Gomes, M. Gomes, Phys. Rev. D85, 085018 (2012), arXiv: 1107.6040.
\bibitem{Hor} P. Horava, Phys. Rev. D {\bf 79}, 084008 (2009), 
arXiv: 0901.3775.
\bibitem{Iengo} R. Iengo, M. Serone, Phys. Rev. D81, 125005 (2010), arXiv: 1003.4430.
\bibitem{our4d} M. Gomes, T. Mariz, J. R. Nascimento, A. Yu. Petrov, J. M. Queiruga, A. J. da Silva, Phys. Rev. D, 92, 065028 (2015), arXiv: 1504.04506. 

%\bibitem{Visser} S. Mukohuyama, Class. Quant. Grav. 27, 223101 (2010), arXiv: 1007.5199; M. Visser, J. Phys. Conf. Ser. 314, 012002 (2011), arXiv: 1103.5587. 


%\bibitem{cpn} S. Das, K. Murthy, Phys. Rev. D80, 065006 (2009), arXiv: 0906.3261.
%\bibitem{ff} A. Dhar, G. Mandal, S. Wadia, Phys. Rev. D80: 105018 (2009),arXiv: 0905.2928.
\bibitem{EP} J. Alexandre, K. Farakos, A. Tsapalis, Phys. Rev. D81, 105029 (2010), arXiv: 1004.4201; C. F. Farias, M. Gomes, J.R. Nascimento, A. Yu. Petrov, A. J. da Silva, Phys. Rev. D85, 127701 (2012), arXiv: 1112.2081; A. M. Lima, J. R. Nascimento, A. Yu. Petrov, R. F. Ribeiro, Phys. Rev. D91, 025027 (2015), arXiv: 1412.2944.
\bibitem{ABJ} J. S. Adler, Phys. Rev. 177, 2426 (1969); J. S. Bell, R. Jackiw, Nuovo Cim. A60, 47 (1969).
\bibitem{JackAmb} R. Jackiw, Int. J. Mod. Phys. B14, 2011 (2000), hep-th/9903044.
\bibitem{DJ} L. Dolan, R. Jackiw, Phys. Rev. D9, 3320 (1974).
\bibitem{highT} C. F. Farias, M. Gomes, J. R. Nascimento, A. Yu. Petrov, A. J. da Silva, Phys. Rev. D89, 025014 (2014), arXiv: 1311.6313.
\bibitem{Das} A. Das, A. Karev, Phys. Rev. D36, 623 (1987).
\bibitem{CSYM}
M. Gomes, J. R. Nascimento, E. Passos, A. Yu. Petrov, A. J. da Silva,
Phys. Rev. D76, 047701 (2007).
\bibitem{Farias} C. F. Farias, J. R. Nascimento, A. Yu. Petrov, Phys. Lett. B719, 196 (2013), arXiv: 1208.3427.
\bibitem{Bakas} I. Bakas, D. Lust, Fortsch. Phys. 59, 937 (2011), arXiv: 1103.5693; I. Bakas, Fortsch. Phys. 60, 224 (2011), arXiv: 1110.1332.
\bibitem{Bilal} A. Bilal, ``Lectures on anomalies'', arXiv: 0802.0634.
\bibitem{Fuji} K. Fujikawa, H. Suzuki, Path Integrals and Quantum Anomalies, Clarendon Press, Oxford, 2004.
%\bibitem{GR}B. de Wit, M. Grisaru, M. Rocek, Phys. Lett. B374, 297 (1996), hep-th/9601115.
\end{thebibliography}
\end{document}